\documentclass[prd,twocolumn,aps,psfig,showpacs,nofootinbib,nobibnotes,
superscriptaddress]{revtex4}

\usepackage{graphicx,natbib}
\usepackage[figuresright]{rotating}
\usepackage{amssymb}
\usepackage{bm}
\usepackage{epsfig}
\usepackage{color}

\newcommand{\be}{\begin{equation}}
\newcommand{\ee}{\end{equation}}

\newcommand{\bea}{\begin{eqnarray}}
\newcommand{\eea}{\end{eqnarray}}
\newcommand{\bean}{\begin{eqnarray*}}
\newcommand{\eean}{\end{eqnarray*}}

\newcommand{\ba}{\begin{eqnarray}}
\newcommand{\ea}{\end{eqnarray}}

\begin{document}

\title{Polarized Gravitational Waves from Cosmological Phase Transitions}

\author{Leonard Kisslinger}
\email{kissling@andrew.cmu.edu}
\affiliation{McWilliams Center for
Cosmology and Department of Physics, Carnegie Mellon University,
5000 Forbes Ave, Pittsburgh, PA 15213, USA}

\author{Tina Kahniashvili}
\email{tinatin@andrew.cmu.edu}
\affiliation{McWilliams Center for
Cosmology and Department of Physics, Carnegie Mellon University,
5000 Forbes Ave, Pittsburgh, PA 15213, USA}
\affiliation{Department of Physics, Laurentian University, Ramsey
Lake Road, Sudbury, ON P3E 2C,Canada}
\affiliation{Abastumani Astrophysical Observatory, Ilia State University,
3-5 Cholokashvili St., 0194 Tbilisi, Georgia}

\date{\today}

\begin{abstract}
 We estimate the degree of circular polarization for the gravitational waves
generated during the electroweak and QCD phase transitions from the kinetic and
magnetic helicity generated by bubble collisions during those cosmological
phase transitions.
\end{abstract}

\maketitle

\section{Introduction}
Gravitational waves (GWs) astronomy opens a new window to study the physical
processes in the very early universe: relic GWs propagate almost
freely throughout the universe expansion, and thus they retain the information
about the physical conditions and physical processes at the moment of their
generation (see for reviews, \cite{Maggiore:2006uy,Hogan:2006va,Buonanno:2007yg}
and references therein). There are various mechanisms that might generate such
 GWs. In the present paper we focus on the generation of GWs during
cosmological electroweak (EW) and Quantum ChromoDynamic (QCD) phase transitions
(PTs) through the turbulent helical sources which can arise and follow the PT
bubble collisions. The GW generation mechanism associated with bubble
collisions during the first
 order PTs has been widely discussed in literature, starting from
 the pioneering works \cite{Kosowsky:1991ua,Kosowsky:1992rz,Kosowsky:1992vn,Kamionkowski:1993fg} and re-addressed later \cite{Allen:1996vm,Gleiser:1998na,Ahonen:1997wh,Apreda:2001us,Grojean:2006bp,Randall:2006py,Caprini:2007xq,Megevand:2008mg,Huber:2008hg,Leitao:2010yw,No:2011fi,Chialva:2010jt,Leitao:2012tx,Hindmarsh:2013xza,Giblin:2014qia,Hindmarsh:2015qta,Schwaller:2015tja,Ashoorioon:2009}

For a cosmological phase transition to produce strong enough turbulent
motions and magnetic fields, which will result in the detectable signal of GWs, they must be first order PTs, with bubble formation and bubble collisions. For the
EWPT, with the standard EW Lagrangian plus a Stop, the supersymmetric partner
of the top quark, called the MSSM EW Lagrangian, the EWPT is first order PT \cite{Henley:2010}. The QCDPT has been shown to be  first
order PT by lattice gauge calculations \cite{Bazavov:2014pvz,Borsanyi:2013bia}.

Both turbulent motions and matic fields can produce relic GWs through
their anisotropic stresses, see Refs.
\cite{Kosowsky:2001xp,Dolgov:2002ra,Nicolis:2003tg,Kahniashvili:2005qi,Caprini:2006jb,Gogoberidze:2007an,Kahniashvili:2008er,
Kahniashvili:2008pe,Caprini:2009fx,Kahniashvili:2009mf,Caprini:2009yp}.
It has been pointed that the GWs generated by magnetic fields can
be detected through Laser Interferometer Space Antenna (LISA)
\cite{Maggiore:1999vm,Apreda:2001tj,Ungarelli:2000jp,Hughes:2007xm,Kahniashvili:2008pf}.
In difference to the GWs sourced solely by PTs bubble collisions, the presence of turbulent
 (kinetic and magnetic sources) increases the detection prospects
\cite{Chongchitnan:2006pe} not only from EWPT but from
QCDPT  too \cite{Roque:2013ufa}.\footnote{GWs from QCDPT are
potentially detectable through pulsar timing, see Ref. \cite{Caprini:2010xv} and references therein.} One of
main goals of European Space Agency (ESA) - NASA planned join mission LISA
\cite{lisa}, was the detection of low frequency GWs (sub-Hz
region).
The new development of this program is the European only ESA mission, so called
 New Gravitational wave Observatory (NGO) - {\it aka} eLISA (evolved LISA)
\cite{elisa}. One of major parts of its science program consists on the direct
detection of GWs from cosmological PTs,
see Refs. \cite{AmaroSeoane:2012km,AmaroSeoane:2012je,Binetruy:2012ze,
Regimbau:2011rp} for details.

In the present paper we extend our previous study Ref.
\cite{Kahniashvili:2009mf}, and we investigate the degree of polarization of
GWs generated via cosmological PTs through helical hydro and magnetized turbulent sources using
the formalism given in Ref. \cite{Kahniashvili:2005qi}. We adjust the previous formalism to
determine the polarization degree of GWs from helical kinetic turbulence to a more complex scenario of MHD turbulence present during the cosmological PTs. More precisely, we use the recent results
of numerical simulations \cite{Tevzadze:2012kk,Kahniashvili:2012uj,Brandenburg2015} to set the statistical properties of helical MHD turbulence.  Another difference from the formalism of Ref. \cite{Kahniashvili:2005qi} consists in computing the energy density and peak frequency of GWs using the analogy with acoustic waves production by hydrodynamical turbulence \cite{Gogoberidze:2007an} (which we can call {\it aeroacoustic} approach \cite{aero1,aero2,aero3}).

Charge-conjugation-Parity (CP) violation is necessary for the production
of magnetic helicity via bubble collisions, \cite{Kisslinger:2002mu}.
EWPT and QCDPT bubble collisions result in development of
helical (kinetic or/and magnetic) turbulence, due in part to CP violation,
which will lead to circularly polarized GWs background. In the case of
strong enough helical sources \cite{Kahniashvili:2008er,Kahniashvili:2008pe}, the degree of polarization is potentially detectable
\cite{Nishizawa:2011eq,Seto:2006hf,Seto:2006dz,Seto:2007tn,Seto:2008sr,
Vallisneri:2008ye,Nishizawa:2009jh}.\footnote{The indirect tool to detect
circularly polarized GWs consists on searching parity violating signals on
cosmic microwave background maps, see Refs \cite{Lue:1998mq,Caprini:2003vc,
Gluscevic:2010vv,Saito:2007kt} for original studies and Ref. \cite{Ade:2013nlj}
for a review.}

In our present study
we follow the helical (chiral) magnetic fields generation scenarios (during PTs through bubble collisions) presented in Refs.
\cite{Stevens:2007ep,Stevens:2009ty}
(EWPT) and Refs. \cite{Kisslinger:2002mu,Forbes:2000gr} (QCDPT) (see also Ref.
\cite{Kahniashvili:2009mf} for a brief review of these models). Upon
generation the magnetic field starts to interact with primordial plasma that
leads to development of magnetically dominant MHD and secondary kinetic
turbulence, for pioneering studies see Refs. \cite{Christensson:2000sp,Christensson:2002xu,
Son:1998my,Muller:2000zz}. In what follow we adopt the results of numerical
simulations of Refs. \cite{Tevzadze:2012kk,Kahniashvili:2012uj,Brandenburg2015},
\footnote{We underline the nature of the secondary character of fluid motions,
because the bubble collision itself might lead to development of purely
hydrodynamical turbulence  during PTs, see Refs.
\cite{Nicolis:2003tg}, while here we note that bubble collisions result in
generation of  magnetic fields
 \cite{Stevens:2007ep,Stevens:2009ty,Kisslinger:2002mu,Forbes:2000gr}.} and their phenomenological
 interpretation given in
Refs. \cite{Banerjee:2004df,Campanelli:2007tc}

The structure of the paper is as follows: In Sec. II we review the GWs
generation formalism and define the circular polarization degree of GWs. We
discuss the hydro and MHD helical turbulence modeling in Sec. III and compute
the GW signal and its polarization  in Sec. IV. We give our results for both
EWPT and QCDPT generated GWs in Sec. IV,
and we conclude in Sec. V. We use natural ($\hbar = 1 = c$)
Lorentz-Heaviside units.

\section{Gravitational Waves Generation Overview}
We assume that GWs are generated through kinetic and MHD turbulence which
follow the PT bubble collisions \cite{Kamionkowski:1993fg,Kosowsky:2001xp,Dolgov:2002ra,Nicolis:2003tg,Espinosa:2010hh}. To be general as possible we present the
common description for EWPT and QCDPT, defining the PT temperature as $T_\star$
($T_\star=100$ GeV for EWPT and $T_\star=0.15$ Gev for QCDPT, and the typical
{\it proper}
length scale through $l_0$ which can be associated with the PT bubble size
$l_b$ (the assumption $l_0 \simeq l_b$ is well justified because in our
theory bubble collisions during PT generate a magnetic field at bubble walls,
and this initial field starts to interact with primordial plasma resulting in
development of kinetic and MHD turbulence with a typical length scale that
corresponds to the magnetic field injection scale) \cite{Kahniashvili:2009qi}.
 We also define $g_\star$ as a number of relativistic degrees of freedom: for
the standard model we have $g_*= 106.75$ as $T \rightarrow \infty$.
($g_\star=100$ for EWPT and $g_\star=15$ for QCDPT). In our further consideration
we assume that the
duration of the turbulent sources, $\tau_T$ are short compared to the universe
expansion time-scale at PTs, i.e. $\tau_T \leq H_\star^{-1}$ with $H_\star^{-1}$
the Hubble radius at PT. This assumption makes possible to neglect the
expansion of the universe, although limits our consideration by the GWs
background generation {\it only} from PT, and completely neglects
GWs arising from decaying turbulence (which might last log-enough after the end
of PTs).

GWs (the tensor metric perturbations above the standard
Friedmann-Lema\^\i tre-Robertson-Walker homogeneous and isotropic background)
are generated from turbulence (including kinetic and magnetic fluctuations)
through the presence of anisotropic stresses as
\begin{equation}
\nabla^2 h_{ij}({\mathbf x}, t) - \frac{\partial^2}{\partial t^2}
h_{ij}({\mathbf x}, t) = -16\pi G \Pi_{ij}^{(T)}({\mathbf x}, t),
\label{eq:01}
\end{equation}
where   $h_{ij}({\mathbf x}, t)$ is the tensor metric perturbation, $t$ is
physical time, $i$ and $j$ are spatial indices (repeated indices are summed),
and $G$ is the gravitational constant. We have neglected the term
$\propto  \partial h_{ij}({\mathbf x}, t)/\partial t$ due to our assumption of
the short duration of turbulence. $P_{ij}^{(T)}$ (the script "$T$"
indicates that we are interested on the tensor part of the turbulent source)
 is the traceless part of the stress-energy tensor $T_{ij}({\mathbf x}, t)$,
 which is constructed from kinetic (K) or  magnetic (M)
turbulence normalized vector fields\footnote{The kinetic and magnetic perturbation
stress-energy tensor are
\begin{eqnarray}
T_{ij}^{(K)} ({\mathbf x},t) = {\rm w} u_i({\mathbf x},t) u_j({\mathbf
x}, t),
\label{tensorK}
\\
T_{ij}^{(M)} ({\mathbf x},t) = {\rm w} b_i({\mathbf x},t) b_j({\mathbf
x}, t),
\label{tensorM}
\end{eqnarray}
where ${\rm w} = \rho+p$ is enthalpy of the fluid with  density energy, $\rho$,
and pressure $p$,  ${\bf u}({\mathbf x},t)$ is the kinetic motion velocity
field and ${\bf b}({\mathbf x},t)$ is normalized magnetic field,
${\bf b}={\bf B}/\sqrt{4\pi {\rm w}}$, that represents the Alfv\'en velocity,
$v_A$ of the magnetic field.
The normalized energy of the magnetic field is then ${\mathcal E}_M (\eta) =
\langle {\bf b}^2(t)\rangle/2$, while the normalized kinetic energy density is
given through ${\mathcal E}_K(t) = \langle {\bf u}^2(t)\rangle/2$. The advantage
 of such representation consists on eliminating the expansion of the universe,
since physical and comoving values of the normalized magnetic field amplitude
are the same.    }
(as we will show below the equipartition is established between kinetic and
magnetic turbulent motions which simply doubles the source term, i.e.
$\Pi_{ij}^{(K)}+\Pi_{ij}^{(M)} \simeq 2\Pi_{ij}^{(T)}$) given by \cite{W}:
\begin{equation}
\Pi_{ij}^{(T)}({\mathbf x}, t) = T_{ij}({\mathbf x}, t) - \frac{1}{3} \,
\delta_{ij} T({\mathbf x}, t), \label{eq:02}
\end{equation}
here $T \equiv [T]_k^k$ is the trace of the $T_{ij}$ tensor.

As we can expect the kinetic and magnetic turbulent fluctuations
generate stochastic  GWs, which can be characterized by the wave number-space
two-point  function as,
\begin{eqnarray}
\langle h^{\star}_{ij}({\mathbf k},t) h_{lm} ({\mathbf k'},t+\tau)\rangle
=(2\pi)^3 \delta^{(3)}({\bf k}-{\bf k'}) \times ~~~~~~~~~~
\nonumber\\
\times \left[ {\mathcal
M}_{ijlm} ({\mathbf{\hat k}}) H(k,\tau)
+ i{\mathcal A}_{ijlm}({\mathbf{\hat k}}) {\mathcal H} (k,\tau) \right].~
\label{gw1}
\end{eqnarray}
Here we use the Fourier transform pair of the tensor perturbation as:
$h_{ij}({\mathbf k},t)=\int d^3\!x \,
   e^{i{\mathbf k}\cdot {\mathbf x}} h_{ij}({\mathbf x},t)$
and $h_{ij}({\mathbf x},t)=\int d^3\!k \,
   e^{-i{\mathbf k}\cdot {\mathbf x}} h_{ij}({\mathbf k},t)/(2\pi)^3$. The brackets $\langle ... \rangle$ denote an ensemble average over realization of
the stochastic source.
The spectral functions $H ({k}, t)$ and ${\mathcal H}({k}, t)$ determine
the GW amplitude and polarization,
 $4 {\mathcal M}_{ijlm} ({\mathbf{\hat k}}) \equiv
P_{il}({\mathbf{\hat k}})P_{jm}({\mathbf{\hat k}})+P_{im}({\mathbf{\hat k}})
P_{jl}({\mathbf{\hat k}})-P_{ij}({\mathbf{\hat k}})P_{lm}({\mathbf{\hat k}})$, and
 $8 {\mathcal A}_{ijlm}({\mathbf{\hat k}})
\equiv {\hat {\bf k}}_q \big[P_{jm} ({\mathbf{\hat k}})\epsilon_{ilq} +
P_{il}({\mathbf{\hat k}})
\epsilon_{jmq} + P_{im} ({\mathbf{\hat k}})\epsilon_{jlq} +
P_{jl}({\mathbf{\hat k}}) \epsilon_{imq} \big]$ are
tensors, with the projection tensor
$ P_{ij}({\bf\hat k}) = \delta_{ij}-{\hat k}_i{\hat k}_j$ (with $\delta_{ij}$ -
the Kronecker delta, ${\hat k}_i=k_i/k$ and $k=|{\bf k}|$),  $\epsilon_{ijl}$
is is the totally antisymmetric symbol.
 We choose GW propagation direction pointing the unit vector
 ${\hat {\mathbf e}}_3$, and we use the usual circular polarization basis
tensors  $e^{\pm}_{ij} = -({\bf e}_1
\pm i{\bf e}_2)_i \times ({\bf e}_1 \pm i {\bf e}_2)_j/\sqrt{2}$. We
define two states
$h^+$ and $h^-$ corresponding  to right- and left-handed
 circularly  polarized
GWs,  $h_{ij}=h^+e^+_{ij} + h^-
e^-_{ij}$.
Through above notations the circular polarization degree was derived
for GWs from Gamma Ray Bursts by \cite{Meszaros:2003vw} and in the context of
cosmological GWs is reproduced \cite{Kahniashvili:2005qi},
\begin{equation}
{\mathcal P}^{\rm GW}(k) =  \frac {\langle h^{+ \star}({\mathbf k})
h^{+}({\mathbf k'}) -
 h^{- \star}({\mathbf k},) h^{-}({\mathbf k'}) \rangle}
{\langle h^{+ \star}({\mathbf k}) h^{+}({\mathbf k'}) +
 h^{- \star}({\mathbf k}) h^{-}({\mathbf k'}) \rangle}
=\frac{{\mathcal H}(k)}{H(k)}.~ \label{degree}
\end{equation}
Here we omit the time dependence of the polarization degree $ {\mathcal P}^{\rm GW}(k)$.

As we already underlined we are interested on GWs generation only from a short
duration sources acting during PTs. After generation the GWs propagate almost
freely, and we account for the expansion of the universe
by a simple re-scaling of the frequency and the amplitude by a factor
equal to
\begin{equation}
\frac{a_\star}{a_0} \simeq 8 \times 10^{-16} \left(\frac{100\,{\rm
GeV}}{T_\star}\right) \left(\frac{100}{g_\star}\right)^{{1}/{3}},
\label{a-ratio}
\end{equation}
This factor is safely canceled when computing ${\mathcal P}^{\rm GW}(k)$, although the
more complex consideration of decaying turbulence (long lasting sources) will
make  ${\mathcal P}^{\rm GW}(k)$ time dependent function.

To estimate the polarization degree of GWs from PT generated helical fields we
need to compute two spectral functions ${\mathcal H}(k)$ and $H(k)$
at the moment of PT, which are determined by the helical anisotropic sources
$\Pi_{ij}^{(K)}$ and $\Pi_{ij}^{(M)}$. In our previous work we have computed the
typical amplitude and
frequency of GWs generated during cosmological PTs \cite{Kahniashvili:2009mf}.
In Ref.
\cite{Kahniashvili:2005qi} the polarization degree of GWs from kinetic (hydro)
 turbulence has been estimated. It has been shown that fully helical turbulence
 leads to ${\mathcal P}^{\rm GW}(k) \rightarrow 1 $. In the present work we follow the
GWs generation formalism from helical magnetized sources presented in Refs.
\cite{Kahniashvili:2008er,Kahniashvili:2008pe}, and apply it to the EWPT and
QCDPT cases.

\section{Kinetic and MHD Turbulence Modeling}

The magnetic field  amplitude (i.e. total magnetic field energy density) is
strongly limited by
the big bang nucleosynthesis (BBN) bound requesting that the total magnetic field
energy density cannot exceed the 10$\%$ of the radiation density at the moment
of the magnetic field generation. In terms of the {\it effective } comoving
magnetic
field value, $B_{\rm eff} \simeq 8.4 \cdot 10^{-7} (100/g_\star)^{1/6}$ Gauss (G) or
in the terms of Alfve\'n velocity $v_A \equiv |{\bf b}| \leq 0.4$
\cite{Kahniashvili:2009qi}.

As we noted above the magnetic field generated at one scale (the magnetic field
initial
spectrum can be approximated being peaked at the typical wavenumber
$k_0=2\pi/l_0$, so in Fourier ${\bf k}$-space described by
$\delta^{(3)}({\bf k}-{\bf k}_0)$ function) after interactions
with plasma leads to development of turbulence,\footnote{Primordial plasma is a perfect conductor with extremely high
values of kinetic and magnetic Reynolds numbers, and current numerical simulations
are still behind to approach necessary resolutions and timescales to describe adequately physical conditions and processes in the early universe.
}
and  the sharply peaked initial
spectrum is redistributed
respectively. For isotropic {\it stationary} turbulence the  normalized magnetic
vector field two-point correlation function is:
\begin{equation}
\langle b_i^* ({\bf k}) b_j({\bf k'}) \rangle = (2\pi)^3
\delta^{(3)}({\bf k} -{\bf k'}) \,  F_{ij}^M\!({\bf k})
  \label{2-point}
\end{equation}
where
\begin{equation}
F_{ij}^M\!({\bf k}) =  P_{ij}({\bf {\hat k}}) S_M(k) + i \epsilon_{ijl}
{{\hat k}_l} A_M(k).
\label{F-spectrum}
\end{equation}
The power-law spectral function $S_M(k) = S_0 k^{n_S}$ and $A_M(k)=
A_{0}k_0^{n_S-n_A}k^{n_A}$ determine the energy density and
current helicity of the magnetic field\footnote{Magnetic helicity defined as
$\langle {\bf A} ({\bf x})\cdot {\bf B}({\bf x})
\rangle $ is a gauge-dependent quantity, while  normalized (or regular
expressed through ${\bf B}({\bf x})$) current helicity $\langle {\bf b}
({\bf x})\cdot
 [{\mathbf \nabla} \times {\bf b}({\bf x})]  \rangle$ is gauge independent, see
Ref.
 \cite{Kunze:2011bp} for details, also it allows the direct analogy with the
kinetic helicity $\langle  {\bf u}({\bf x}) \cdot [{\mathbf \nabla} \times
{\bf u}({\bf x})]
 \rangle$, and thus to consider both helical sources in a common formalism. }
 and $n_S$ and $n_A$ are magnetic field and helicity spectral indices which
determine the spatial distribution of the magnetic field and its helicity.
 The establishment of stationary turbulence with the stationary (time-
independent) two-point correlation function given through Eqs. (\ref{2-point}) -
  (\ref{F-spectrum}) requires the presence of long-lasting sources. To account
for the short acting PT turbulent source (the turbulence duration time
$\tau_{\rm T}$ is short enough compared to the universe expansion time-scale
$H_\star^{-1}$) we have to modify the spectra $S_M(k)$ and $A_M(k)$ making them
time dependent, see below.

Following the description of hydro- and MHD turbulence generated during PTs, we
distinguish three spatial spectral sub-regimes
of turbulent fluctuations: (i) the large scale decay range $k_{H_\star}<k<k_0$
(where physical wavenumbers $k_{H_\star}=2\pi/H_\star^{-1}$ and $k_0$
correspond to the PT Hubble length scale and the largest PT length size); the
minimal wavenumber corresponds to the Hubble scale $H_\star^{-1}$ beyond which
causally generated magnetic field is {\it frozen-in} and any interactions are
forbidden due to causality requirement; (ii) the turbulent (or  so called
{\it inertial}) range  $k_0<k<k_D$ (where $k_D$ the damping scale of turbulence
 through viscous dissipation and magnetic resistivity, which is determined by
plasma properties); (iii) the damping range $k>k_D$. All these typical
wavenumbers ($k_{H_\star}$, $k_0$, and $k_D$) are time-dependent due to
interactions of magnetic field with plasma and the expansion of the universe.
Magnetic helicity presence plays here a crucial role leading to re-arrangement
of the helical structure at large scales \cite{B03}. The expansion of the
universe lead to additional effects: namely, the PT bubble size determined
length scale $l_0$ is strengthened by a factor $a(t)/a_\star$ (being
 $\propto t^{1/2}$ during the  radiation-dominated epoch and $\propto t^{2/3}$
during the matter-dominated epoch),  while the Hubble length scale
$H_\star^{-1} \propto t$. As a result a perturbation with $k_{H_0}<k<k_{H_\star}$
(with $k_{H_0}=2\pi/H_0$ and $H_0$ is the today Hubble radius) will enter the
Horizon at some point \cite{Tashiro:2005hc}. Since we are focused on the short
duration sources, we will completely neglect the GWs signal for the large-scale
decay range $k<k_0$ (where the spectral shape of the field is given through the
causal Batchelor spectrum with $n_S=2$ \cite{Durrer:2003ja}). Obviously the GWs
signal from the viscous damping range $k>k_D$ is also negligibly small.

The realizability condition implies that $|A_M(k)| \leq S_M(k)$ (the modulus
sign reflects a possibility of having positive or negative helicity). The
spectral indices values $n_S$ and $n_A$ strongly depend on the turbulence
model. In the inertial range ($k_0<k<k_D$), for non-helical turbulence the
Kolmogoroff model implies $n_S=-11/3$. Some models lead to the different
spectral shapes such as $n_S=-7/2$ - Iroshnikov-Kraichnan model for magnetized
turbulence \cite{IK1,IK2}, $n_S=-4$ - the weak turbulence model
\cite{Brandenburg:2009tf} or magnetically dominant turbulence
\cite{Brandenburg:2014mwa}. In the presence of helicity the consideration is
even more complex, and requires careful investigation through numerical
simulations that it is beyond the scope of the present paper. Based on the
phenomenological dimensionless description, if the process is driven by the
magnetic energy dissipation at small scales, it is assumed $n_S=-11/3$ and
$n_A=-14/3$ (so called {\it helical} Kolmogoroff model)  \cite{my75}, while if
the process is determined by helicity transfer (inverse cascade) and helicity
dissipation at small scales it is adopted $n_S=-13/3=n_A$ \cite{moiseev}.

To account the short-duration turbulence (not enough to establish the stationary
 turbulent motions) we have to consider turbulent fluctuations
time-de-correlation which can be accounted for via introducing the
characteristic function $f(\eta(k), \tau) $ (with $\eta(k)$ the autocorrelation
function),
\cite{kraichnan}:
\begin{equation}
f(\eta(k), \tau) = exp \big[-\frac{2\pi^2}{9} \big(\frac{\tau}{\tau_0} \big)
 {K}^{4/3} \big]
\label{f-decorr}
\end{equation}
with $\tau_0$ - the largest turbulent eddy turn-over time, $K
\equiv k/k_0$. In this case, to determine the magnetic field two point
correlation function in real (${\bf x}$) space, we have to account for magnetic
field fluctuations at {\it different} time moments and at {\it different}
positions, i.e. $\langle b_i ({\bf x}, t) b_j({\bf x}+{\bf R}, t+\tau) \rangle$.
Accordingly, in Fourier space, the two-point correlation function will be
determined by the  ${\bar F}_{ij}^M\!({\bf k}, t)$  with the time dependent
spectral functions $S_M(k,t)$ and $A_M(k,t)$:
\begin{eqnarray}
 {\bar F}_{ij}^M\!({\bf k}, t) &=& \Big [P_{ij}({\bf {\hat k}}) S_M(k,t) +
i \epsilon_{ijl}{{\hat k}_l} A_M(k,t) \Big] \nonumber
\\ &\times & f(\eta(k), t)
 \label{f2a}
 \end{eqnarray}
Comparing with the stationary spectrum, see Eq. (\ref{2-point}), we see that
formally we replace
$ F_{ij}^M\!({\bf k}) \equiv F_{ij}^M\!({\bf k}, t)$,
by ${\bar F}_{ij}^M\!({\bf k}, t) = F_{ij}^M\!({\bf k}, t)f(\eta(k), \tau)$,
To avoid a complex description of accounting time
dependence of  $S(k,t)$ and $A(k,t)$, we use the Proudman argument for
kinetic turbulence
\cite{aero1}, according which {\it the description of decaying turbulence
lasting for $\tau_T$ can be replaced by the description of stationary
turbulence with time duration of $\tau_T/2$}. Below we briefly discuss our
approach.

Turbulence during PTs generated through magnetic helicity  can described
through two major stages \cite{Kahniashvili:2008pe}: during the first stage the
main process is determined by the magnetic energy direct cascade that last few
largest eddy turnover times $\tau_0=2\pi/(k_0 v_0)$, where $v_0 <1 $ is the
turbulent eddy velocity ($v_0 \simeq M$ for kinetic turbulence where $M$ is
the Mach number  and $v_0 \simeq v_A$ for magnetic turbulence) determined by PT
and magnetogenesis model parameters, see Ref. \cite{Nicolis:2003tg,Leitao:2014pda}, i.e.
$\tau_{\rm T} = s_0 \tau_0$ (with $s_0 = 3 -5$). The magnetic field induces
vorticity fluctuations, and at the end of the first (semi)equipartition between
kinetic and magnetic energies is reached, that results in doubling the value of
the source for GWs. The magnetic energy density power spectra
are then determined by the proper dissipation rate per unit enthalpy
$\varepsilon_M$ as: $S_0 = \pi^2 C_K \varepsilon^{2/3}$ with $C_K$ constant
order of unity, and $\varepsilon = k_0v_0^3$. Note that the autocorrelation
function $\eta(k) = \varepsilon^{1/3} k^{2/3}/\sqrt{2\pi}$ \cite{Terry:2007at}.
Although the Kolmogoroff model is valid {\it only} for non-relativistic
turbulence, while during PTs we might deal with $v_0 \simeq 1$ (relativistic
turbulence), our estimates for the amplitude and polarization degrees of GWs
signal are qualitatively justified, see \cite{Kosowsky:2001xp}.
The second stage consists on helicity transfer (inverse cascade). The scaling
laws for this stage are still under debate. Based on our previous consideration
\cite{Kahniashvili:2008pe}, we assume that (i) the details of the scaling laws
during this stage will not affect substantially our estimates; (ii) instead of
considering decay-turbulence we will again consider the stationary turbulence
with scale-dependent duration time. Then we obtain  for the helical Kolmogoroff
model, $A_0 = \pi^2 C_K \sigma /(k_0 \varepsilon^{1/3})$, where $\sigma$ is the magnetic helicity dissipation rate per unit enthalpy, leading to $A_0/S_0 =
\sigma/(\varepsilon k_0)$. The helical Kolmogoroff model  is
mainly relevant for weakly helical fields, $|A(k)| \ll S(k)$, which is a case of
 magnetic fields generated during PTs.\footnote{Note that the
substantially helical case is usually determined by the helicity transfer
(inverse cascade), $S_0 = C_S \sigma^{2/3} $ and $A_0 = C_A \sigma^{2/3}$
\cite{moiseev}.}
According to results of recent numerical simulations, see Ref. \cite{Brandenburg2015},
the weakly
helical turbulence even accounting for the free decay of turbulence, show an
establishment of the spectra in a good agreement with the helical Kolmogoroff
model, as well as equipartition between magnetic and kinetic energy densities.
Thus we adopt $n_S=-11/3$ and $n_H=-14/3$ for the inertial range, with
$S_0 = \pi^2 k_0^{2/3} v_0^2$ and $A_0=h S_0$ with $h$ which determines the fraction of helicity dissipation,
$h \equiv \sigma/(\varepsilon k_0)$.  The turbulence fluctuation velocity $v_0=v_A$ is determined by the
magnetogenesis mechanism and for the model of our interest is given by $v_0
\simeq 0.2 $ ($B_{\rm eff, in } \simeq 5 \cdot 10^{-7}$ G)
\cite{Kahniashvili:2009mf} for EWPT model of Refs.
\cite{Stevens:2007ep,Stevens:2009ty} and $v_0 \simeq 0.01$ ($B_{\rm eff, in }
 \simeq 2 \cdot 10^{-8}$ G) \cite{Tevzadze:2012kk} for QCDPT model of Refs.
\cite{Kisslinger:2002mu,Forbes:2000gr}.

\section{Gravitational Waves Signal Amplitude and Polarization}
In this section we compute the GW signal (stain amplitude) and polarization
degree from the hydro and MHD turbulence generated during the first order
cosmological PTs. To determine the amplitude of GWs we proceed as it is
described in Ref. \cite{Kahniashvili:2008pe}. We derive the energy
density spectrum of the GWs at the
end of PT (in our approximation the end of turbulence).
The energy density of GWs is given through the ensemble average as
\begin{equation}
\rho_{GW}({\bf x},t)= \frac{1}{32\pi G} \langle
\partial_t h_{ij}({\mathbf x},t) \partial_t h_{ij} ({\mathbf x},t)\rangle , \label{energy-density}
\end{equation}
As we noted the rescaling of the GWs amplitude and frequency given through
Eq. (\ref{a-ratio}) is irrelevant when computing the polarization degree, while
it is crucial for estimation of GWs energy density.

\subsection{Gravitational Wave Signal}

Assuming the homogeneous and isotropic turbulent source lasting for $\tau_T$,
and using far field approximation, see \cite{Gogoberidze:2007an}, the total
energy density of GWs at a given spatial point and a given time
can be obtained by integrating over all sources within a spherical
shell centered at that observer, with a shell thickness
corresponding to the duration of the turbulent source (in our case the duration
of  PT), and a
radius equal to the proper distance along any light-like path
from the observer to the source (causality requirement), and then
\begin{equation}
\rho_{GW}(\omega) = \frac{d\rho_{GW}}{d\ln\omega}= 16\pi^3\omega^3
G \, {\rm w}^2 \tau_T H_{ijij}(\omega,\omega), \label{GW-spectrum}
\end{equation}
where $\omega$ is the angular frequency measured at the moment of
generation of GWs, and $H_{ijij}(\omega,\omega)$ is a
complicated function of $\omega$ (which is computed through using of
aero-acoustic approximation and Millionshchikov quasi-normality \cite{my75}),
given as
\begin{eqnarray}
H_{ijij}({\bf k}, \omega) \simeq  H_{ijij}(0,\omega) = \frac{7
C_K^2 \varepsilon}{6 \pi^{3/2}} \int_{k_0}^{k_D} \!
\frac{dk}{k^6} \times
\nonumber \\
\exp\!\left( -\frac{\omega^2}{\varepsilon^{2/3}
k^{4/3}} \right)\!{\rm erfc} \!\left(
-\frac{\omega}{\varepsilon^{1/3} k^{2/3}} \right)
\label{H-ijij}
\end{eqnarray}
Here, ${\rm erfc}(x)$ is the complementary error function defined
as $\mbox{erfc}(x) = 1 - \mbox{erf}(x)$, where $\mbox{erf}(x) =
\int_0^x dy \exp(-y^2)$ is the error function~\cite{Gradshteyn}.
The integral in Eq.~(\ref{H-ijij}) is dominated by the large
scale ($k \simeq k_0$) contribution so, for  direct-cascade
turbulence during the first stage (direct cascade, see Sec. III), the peak
frequency is
\begin{equation}
\label{OmegaMaxI} \omega_{\rm max}^{(I)} \simeq k_0 M.
\end{equation}
where $M$ is Mach number.
To compute the GWs signal arising from the inverse cascade  stage
we have consider two models separately: Model A assumes that the correlation
length during the inverse cascade scales as $\xi_M \propto t^{1/2}$ and Model B
corresponds to the correlation length scaling as $\xi_M \propto t^{2/3}$. We
obtain that in both models peak frequency during the second stage are equal and
are determined by the Hubble frequency as \cite{Kahniashvili:2008pe}:
\begin{equation}
\label{OmegaMaxBM} \omega_{\rm max}^{(II)} \simeq
 2\pi H_\star \,
\end{equation}
while the GW amplitude are slightly different in Model A and B as
\begin{eqnarray}
H_{ijij}^{(\rm A)} ({\bf k}, \omega) \simeq H_{ijij}(0,\omega) = \frac{7
C_1^2 M^3 \zeta_\star^{3/2}\!}{12 \pi^{3/2} k_0}
 \int_{k_{S}}^{k_0}
\! \frac{dk}{k^4} \nonumber \\  \exp\!\left( -\frac{\omega^2 k_0^2}{\zeta_\star
M^2 k^4} \right)\!{\rm erfc} \!\left( - \frac{\omega
k_0}{\zeta_\star^{1/2} M k^2} \right) \!, \label{modelA}
\end{eqnarray}
and
\begin{eqnarray}
H_{ijij}({\bf k}, \omega)^{(\rm A)}  \simeq H_{ijij}(0,\omega) = \frac{7
C_1^2 M^3 \zeta_\star^{3/2}\!}{6\pi^{3/2} k_0^{3/2}}
\int_{k_S}^{k_0} \! \frac{dk}{k^{7/2}}
\nonumber \\
\exp\!\left(
-\frac{\omega^2 k_0}{\zeta_\star M^2 k^3} \right)\!{\rm erfc}
\!\left( - \frac{\omega k_0^{1/2}}{\zeta_\star^{1/2} M k^{3/2}}
\right) \!. \label{modelB}
\end{eqnarray}
Here $\zeta_\star$ determines the amount of initial magnetic helicity  and is
equal to $\zeta_\star = \langle {\bf a}({\bf x}) \cdot {\bf b}({\bf x}) \rangle
/(\xi_M {\mathcal E}_M) $ (with ${\bf a} ({\bf x}) = {\bf A} ({\bf x})/{\rm w}$
normalized vector potential), and  $k_S=2\pi/l_S$ is the typical scale at which
the inverse cascade stops: either because the cascade time $\tau_{\rm cas} $
reaches the expansion time scale $H_\star^{-1}$ or because the characteristic
length scale $l_S \simeq \xi_M$ reaches the Hubble radius $H_\star^{-1}$. The
value of $k_S$ can be found by using above conditions, being equal to $k_S = k_0
 \zeta_\star^{-1/4} (\gamma/M)^{1/2}$. Note that the integrals in expressions
Eqs. (\ref{modelA}) - (\ref{modelB}) are In this
dominated by the  large scale $k \simeq
k_S$ contributions and are maximal at
$\omega_{\rm max}^{(II)}$.

\subsection{Gravitational Wave Polarization}
To compute the polarization degree of GWs we need to estimate the tensor
perturbations source two point correlation function ${\mathcal F}_{ijlm} (k,
\tau) \equiv \langle \Pi^{(T) \star}_{ij} ({\mathbf k}, t) \Pi^{(T)}_{lm}
({\mathbf k^\prime},t^\prime +\tau )\rangle $, which can be expressed through the
forms ${\mathcal M}_{ijlm}$ and ${\mathcal A}_{ijlm}$ (which are defined below
Eq.~(\ref{gw1})), as
\begin{eqnarray}
{\mathcal F}_{ijlm} (k, \tau) & = &(2\pi)^3 \delta^{(3)}({\bf k}-{\bf k'})
\nonumber\\
& \times & \left[ {\mathcal M}_{ijlm} {\mathcal S}(k,\tau)
+ i{\mathcal A}_{ijlm} {\mathcal Q } (k,\tau) \right].
\label{pi}
\end{eqnarray}
As we discussed above for the non-stationary turbulence $S(k, \tau)$ and
$A(k, \tau)$ are complex functions of $\tau$ and $k$ (see Eqs. (\ref{f-decorr})
- (\ref{f2a}) for the time
de-correlation function and the magnetic field spectrum). Following
Ref. \cite{Kahniashvili:2005qi} we split the spatial and temporal dependence
as  $S(k, \tau) = S(k) D_S (\tau) $ and $A(k, \tau)= A(k) D_A(\tau)$ which is
a valid approximation for $k \simeq k_0$ (the range which mostly contributes
to the GWs signal). Generalizing the stationary case \cite{Caprini:2003vc}
through accounting for the time-dependent functions $D_S(\tau)$ and $D_A(\tau)$
the forms for ${\mathcal S}(k,\tau)$ and ${\mathcal Q}(k,\tau)$ are given as,
\begin{eqnarray}
{\mathcal S}(k, \tau)&=&
\frac{{\rm w}^2}{(2\pi)^6}
\!\int\!d^3p_1\!
\int\!d^3p_2 \delta^{(3)}({\bf k}-{\bf p_1}-{\bf p_2})
\nonumber \\
&&\times \left[(1+\alpha^2)(1+\beta^2)D_S^2(\tau)S(p_1)S(p_2)
+\right.
\nonumber \\
&& \left.
~~~~~~~~~~+4\alpha \beta D_A^2(\tau)
A(p_1)A(p_2)\right], \label{tensor-source-sym}
\\
{\mathcal Q}(k,\tau) &= &\frac{{\rm w}^2 D_S(\tau)D_A(\tau)}{128\pi^6}
\int\!d^3p_1\!\int\!d^3p_2\,
\nonumber\\ &\times &
\delta^{(3)}({\bf k}-{\bf p_1}-{\bf p_2})
\left[ (1+\alpha^2)\beta S(p_1)S(p_2)
\right.
\nonumber \\
&&~~~~~~~~~~~~~~+
\left.
(1+\beta^2)\alpha A(p_1)A(p_2) \right],
\label{tensor-source-hel}
\end{eqnarray}
where $\alpha={\hat {\bf k}}\cdot{\hat {\bf p}_1}$ and $\beta={\hat
{\bf k}}\cdot{\hat {\bf p}_2}$ with ${\hat {\bf p}_1}={\bf p}_1/p_1$ ($p_1 =
|{\bf p}_1$) and ${\hat {\bf p}_2}={\bf p}_2/p_2$ ($p_1 = |{\bf p}_1$).
The helical source term ${\mathcal Q}(k, \tau)$ vanishes
for turbulence  without helicity.
Since in the helical Kolmogoroff model the time de-correlation is mostly
determined by the energy density dissipation,  thus at first order approximation
we can assume that $D_S(\tau) \simeq D_A(\tau)$ (both functions are
monotonically decreasing functions). By next we should connect $\mathcal S$ and
$\mathcal Q$ with the $H(k, t)$ and ${\mathcal H}(k, t)$ functions, which
determine the GWs polarization degree, see. for details Ref.
\cite{Kahniashvili:2005qi}.

Magnetic helicity generated via bubble wall collisions during the first order
PTs is determined by the corresponding energy scales $\Lambda_{\rm PT}$
($\Lambda_{\rm PT}\simeq $ 100 GeV and 0.15 GeV for EWPT and QCDPT respectively);
 and the PT bubble lengths ($l_b$). In  addition, the bubble wall velocity
substantially affects the turbulent motions development \cite{Espinosa:2010hh}.

In the present paper we focus on magnetic helicity generation mechanisms following Refs. \cite{Forbes:2000gr,Kisslinger:2002mu}. In the framework of these magnetogenesis scenarios magnetic helicity during PTs with the magnetic wall in the $x-y$ plane is given by
\begin{equation}
\label{generalhelicity}
  \mathcal{H}_M = A_z B_z ~~ {\rm or} ~~
   \mathcal{H}_M =  \frac{(B_z)^2}{\Lambda_{\rm PT}} \; ,
\end{equation}
We note that the EWPT energy (mass) scale is approximately equal to Higgs mass, $\Lambda_{\rm EWPT} \simeq M_{\rm Higgs}$. The model parameters such as bubble and wall sizes and wall velocity depend on PT modeling, and are determined in
Refs. \cite{Stevens:2007ep,Stevens:2009ty,Kisslinger:2002mu}. We quote also the physical values of
magnetic field amplitudes as $B^{({\rm EW})}_\star \simeq  6.45 \times 10^4
{\rm GeV}^2$
\cite{Stevens:2007ep,Stevens:2009ty}  and $B^{(\rm{QCD})}_\star \simeq
1.5 \times 10^{-3}{\rm GeV}^2$ \cite{Kisslinger:2002mu}.

The fraction of initial magnetic helicity ($\zeta_\star$) can be
expressed in terms of the magnetic field correlation length (which can be taken
to be equal to $l_b$) and the maximal allowed length scale $H_\star^{-1}$, as
$\zeta_\star \simeq l_b/H_\star^{-1}$.
Following Refs. \cite{Stevens:2007ep,Stevens:2009ty} the normalized magnetic
field generated during the first order
EWPT (at the time moment $\simeq 10^{-11}$ sec) is equal to $v_A \simeq 0.2$, and
assuming around 100 bubble within Hubble length scale, fractional helicity is
 $\zeta_\star^{({\rm EW})} \simeq 0.01$. Comoving magnetic helicity itself is
expressed:
\begin{eqnarray}
\label{BHEWPT}
 {\mathcal H}_{M, \star}^{({\rm EW})} \simeq
 \frac {(B^{({\rm EW})})^2}{125 {\rm GeV}} ,
\end{eqnarray}
Assuming that the magnetic field (with $v_A \simeq 0.01$) is correlated over
the wall
 thickness (the QCD momentum 0.15 Gev) \cite{Kisslinger:2002mu},
results in extremely small magnetic helicity generated during the first order
QCDPT (at time moment $\simeq 10^{-5}$ sec),
\begin{eqnarray}
\label{BHQCDPT}
{\mathcal H}_M^{({\rm QCD})}(t_\star) \simeq
\frac{(B^{(\rm{QCD})})^2}{0.15 {\rm GeV}} \; ,
\end{eqnarray}
which corresponds $\zeta_\star^{({\rm QCD})} \ll 1$.
On the other hand, making the field correlated over the bubble length scale,
leads to the fractional helicity  $\zeta_\star^{({\rm QCD})} \simeq 0.2$

Using  the approximation given above, the
polarization degree of GWs, ${\mathcal P}^{\rm GW}(k)$, for the Kolmogoroff helical
turbulence model can be estimated through (in our simplified description the
time dependence is canceled because of $D_S(\tau) \simeq D_A(\tau)$):
\begin{equation}
{\mathcal P}^{\rm GW}(k) = \frac{{\mathcal H}(k)}{H(k)} = \frac{{\mathcal I}_A({K})}
{{\mathcal I}_S({K})}
\end{equation}
where $K\equiv k/k_0$ is a normalized wavenumber, and
\begin{eqnarray}
{\mathcal I}_S(K) &\simeq &  \int\!dP_1~P_1\!\int\!dP_2~P_2
{\bar \Theta}\left[(1+\alpha_p^2)(1+\beta_p^2)P_1^{n_S}P_2^{n_S}
\right.
\nonumber \\
&& \left.+ 4 h^2 \alpha_p \beta_p P_1^{n_A}P_2^{n_A}\right],~~ \label{tensor-sym1_a}
\\
{\mathcal I}_A(K) &\simeq &2h\int\!
dP_1 ~P_1\!\int\!dP_2~P_2 {\bar \Theta}
\left[(1+\alpha_p^2)\beta_pP_1^{n_S}P_2^{n_A}
\right.
\nonumber \\
&& \left.+ (1+\beta_p^2)\alpha_p P_1^{n_A}P_2^{n_S} \right].~ \label{tensor-hel1_a}
\end{eqnarray}
Here $h$ is the  model parameter (which is related to the helicity fraction, see below, and
 we assume it to be equal to 1,  0.5, 0.1.  $P_1=p_1/k_0$,  $P_2=p_2/k_0$, $\alpha_p = (K^2+P_1^2-P_2^2)/(2KP_1)$, $\beta_p =
(P^2+P_2^2-P_1^2)/(2KP_2)$,  ${\bar \Theta} \equiv \theta(P_1+P_2-K)
\theta(P_1+K-P_2) \theta(P_2+K-P_1)$, and $\theta$ is the Heaviside
 step function which is zero (unity) for negative (positive) argument.
The integration limits ranges from $1$ (we discard the source existence for the
wavenumbers below $k_0$) to $k_D/k_0$.

\begin{figure*}[t!]
\includegraphics[width=0.87\columnwidth]{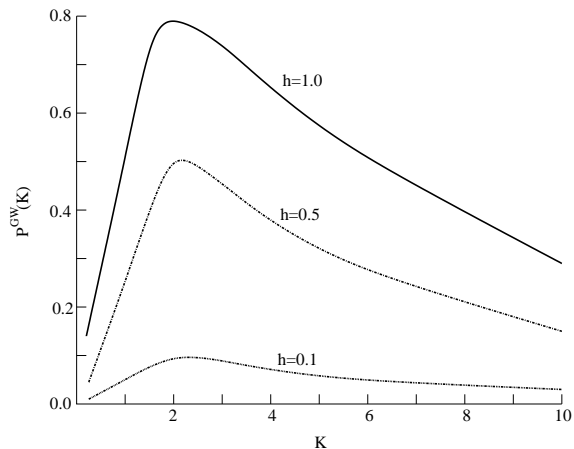}
\caption[]{
The GW polarization degree ${\mathcal P}^{\rm GW}(K)$ ($K=k/k_0$) in terms of the
 model parameter $h=1.0$, $0.5$, $0.1$ }
\label{Figure-1}
\end{figure*}

We emphasize that the fractional helicity parameter $\zeta_\star$ discussed above is defined through normalized magnetic helicity (the integral quantity), while the
parameter $h \equiv \sigma/(k_0 {\varepsilon})$ is defined through the normalized magnetic energy density and normalized magnetic helicity (e.g. it is determined by the power spectra for magnetic energy density and helicity at small length scales). Under the model adopted here (the Kolmogoroff helical turbulence with $n_S=-11/3$ and $n_A=-14//3$) these both quantities coincide $\zeta_\star \simeq h$.

To keep our description as general as possible we present our results for the
GWs polarization degree ${\mathcal P}^{\rm GW}(k)$ in terms of the normalized wavenumber
$K$. As we discussed above the typical wavenumber $k_0$
is determined by the turbulent eddy length scale $l_0$ ($k_0 =2\pi/l_0$) and is significantly
different for EWPT and QCDPT. The model parameter $h$, which determines the
helicity fraction, varies depending on the magnetogenesis model.
The results for ${\mathcal P}^{\rm GW}(k)$ with $h=0.1$, $0.5$, and $1$ are
shown in Figure 1.

\section{Conclusions}
We computed the GWs signal produced during first order
cosmological PTs through hydro and MHD turbulence. We also derive the
polarization degree of GWs assuming the validity of the helical Kolmogoroff
model, shown in Figure 1.
The GWs polarization is present at the background level, and for
maximally helical sources the polarization degree approaches unity at its
maximum, around $k \sim 2 k_0$, and decreases fast at small scales $k \gg k_0$.
The formalism presented in this paper might be used to estimate the
polarization degree of
GWs from helical hydro and MHD turbulence in the differential rotating neutron
stars  \cite{Lasky:2013jfa} or stellar convection \cite{Bennett:2014tba}. Note
from Figure 1 that the detectability of the polarization degree is determined
by the helicity fraction parameter $h$. We plan  in our
future research to make estimates of $h$ values depending on magnetogenesis
models during cosmological PTs.

Previously we have estimated the GWs amplitude, $h_C(f)$, from the first order
EWPT
and QCDPT \cite{Kahniashvili:2009mf}, through assumptions of non-helical magnetic fields   \cite{Stevens:2007ep,Stevens:2009ty,Kisslinger:2002mu}. We have shown that EWPT generated GWs are potentially detectable through LISA-like missions \cite{lisa} in the
case for strong enough EWPT \cite{Kahniashvili:2008pf}
(for QCDPT generated GWs detection prospects see ReF. \cite{Roque:2013ufa}).
 In the present paper we expand our previous results by considering GWs from helical magnetic and hydro turbulence. Probing the circular polarization of GWs background is a challenging task \cite{Nishizawa:2009jh}, and it is quite difficult at the monopole mode. To detect the circular polarization at the dipole or/and  the octopule mode requires at least the system of two unaligned detectors, and LISA was designed ideally to provide detection of these anisotropic components whose magnitudes are small as 1\% of the detector noise  \cite{Seto:2006hf,Seto:2006dz,Seto:2008sr}. The similar detection prospects are expected from eLISA data. The planned eLISA mission \cite{elisa} is originally design to detect un-polarized GWs backgrounds (including GWs from cosmological PTs) \cite{Dufaux:2012rs} with the sensitivity  given on Fig. 1 of Ref. \cite{AmaroSeoane:2012je} (see also comparison with LISA's sensitivity). Although the GWs polarization detection is beyond the currently discussed eLISA science, our study should help further
developments.



\begin{acknowledgments}
It is our pleasure to thank Axel Brandenburg, Leonardo Campanelli, Grigol
Gogoberidze,  and Alexander
Tevzadze for useful discussions and comments. We acknowledge partial support
from the Georgian Shota Rustaveli NSF grant FR/264/6-350/14,
the Swiss NSF SCOPES grant IZ7370-152581, and the NSF
Astrophysics and Astronomy grant AST-1109180.
\end{acknowledgments}

\end{document}